\documentclass{emulateapj} \usepackage{times}

\usepackage{color} 
\usepackage{epsfig}

\newcommand{\be}{\begin{equation}}
\newcommand{\ee}{\end{equation}}
\newcommand{\bea}{\begin{eqnarray}}
\newcommand{\eea}{\end{eqnarray}}

\shorttitle{IR Brightness of Mars}
\shortauthors{Wright}
\journalinfo{}
\submitted{}

\begin{document}

\title{Infrared Brightness Temperature of Mars, 1983-2103}

\author{Edward L. Wright}
\affil{UCLA Dept.\ of Physics \& Astronomy\\
PO Box 951547, Los Angeles CA 90095-1547}
\email{wright@astro.ucla.edu}

\clearpage

\vskip 2.0cm

\begin{abstract}

The predicted infrared brightness temperature of Mars using the 1976 model
of Wright is tabulated here for the period 1983 to 2103.  This model was
developed for far-infrared calibration, and is still being used for JCMT
calibration.

\end{abstract}


\section{Introduction}
\label{sec:intro}

\citet{wri76} improved the far infrared calibration using a time variable brightness
temperature of Mars to replace the constant $T_b = 235$~K that had been used
earlier.  $T_b$ for Mars changes are due both to the eccentricity its orbit
and the large phase angle reached by Mars, which produces a dramatic morning
{it vs.} afternoon temperature change.   \citet{wri80} extended the tabulation
of predicted temperatures for Mars to 1983, and then further predictions 
running until 1995 were provided privately on request.  In 1995 I received 
an E-mail from
Henry Matthews at the James Clerk Maxwell Telescope requesting a further
extension of the tabulated predictions, and an extension was posted on the
WWW\footnote{http://www.astro.ucla.edu/$\sim$wright/old\_mars.txt}.
This table was later extended to 2011 after a further request from the
JCMT in 2001.
As 2011 is approaching, I have extended this tabulation to 2103 and
am posting the tables here.

\section{Calculations}

In 1995 the old code from \cite{wri76} had been lost, and a new version
of the Mars program was created based on the rotating, cratered asteroid
model described in \citet{wri07}.  The modifications involved
\begin{itemize}
\item  setting the number of facets in the craters down to 1,  so the
model is not cratered,
\item setting the number of latitudes to 6, and using Gaussian
quadrature weights and knots for $\sin(\mbox{lat})$,
\item adding a constant 1\% of the noon flux to allow for atmospheric
emission,
\item adding the angle dependent emissivity, $\propto \mu^\beta$,
that was used in \citet{wri76}.
\end{itemize}
While adding the angle dependent emissivity it was noticed that
\citet{wri76} had assumed $\epsilon/(1+\beta)$ for the hemispheric
emissivity instead of the correct $\epsilon/(1+\beta/2)$.
Since the purpose of the calculation was to reproduce the old
Mars model, a temperature fudge factor of $+1.7\%$ in the
sub-solar temperature $T_{SS}$ was
added to the code, along with a $+4.3\%$ increase in the thermal
inertia parameter 
$\Theta = (\kappa\rho C \omega)^{1/2} T_{SS}/[(1-A)F_\odot]$.  
The thermal inertia 
$\sqrt{\kappa\rho  C} = 0.006\;\mbox{cal/cm$^2$/K/s$^{1/2}$}
= 251\;\mbox{J/m$^2$/K/s$^{1/2}$}$,
the emissivity is $\epsilon = 0.9$ and its angular variation is
$\beta = 0.1$, the solar flux at 1 AU is $1367\;\mbox{W/m$^2$}$,
and the albedo is $A = 0.29$.
Replacing the step function
temperature {\it vs.} time model used in \citet{wri76} with the
triangle function used in \citet{wri07} led to an effective hour angle
change of $0.28^\circ$.  With these modifications the new code
gave exactly the same brightness temperature for 82\% of
the entries in the \citet{wri76} and \citet{wri80} tables, when
rounded to the nearest K.  18\% of the entries differed by
$\pm 1$~K.

The positions of Mars and the Earth are computed using the
low precision formulae of \citet{vfl79}.  The Mars pole is
$21^h14^m$, $55^\circ\;02.4^\prime$ (J2000) and the rotation
period is $88775\;\mbox{sec}$.

The modified Julian day MJD in the tables is $\mbox{JD}-2440000$.
A blackbody with no limb darkening at the temperatures given
in the tables will give the same flux as the Mars model at the
tabulated wavelengths, given in $\mu$m:
10, 20, 34, 65, 100, 150, 250 \& 350.

\section{Discussion}

This model is obsolete but these tables could be useful
for understanding far infrared calibrations.
Mars has well known albedo variations, thermal inertia
variations, and craters.  These could be included but then
the model would depend on the local time, and the results
would have to tabulated hourly instead of every 40 days.
The treatment of the atmosphere and especially dust in
the atmosphere is minimal in this model.
During dust
storms the coupling between the atmosphere and the surface
by infrared emission will get much larger than the 1\% assumed
here, and will lead to a larger effective thermal inertia during dust
storms.
Certainly by the end of the time range in these tables there should
be much more information available about thermal conditions on
Mars.  A comparison of this new data to the 1976 Mars model 
will allow one to determine appropriate calibration corrections
for far-infrared and sub-millimeter data.

The code used to calculate the Mars model and generate the 
tables presented here has been appended to the LaTex
file after the \verb"\end{document}".  This paper will be a
Web-only manuscript, available at http://arXiv.org and its mirrors.

\begin{table}[p]
\caption{Brightness Temperature of Mars:1983-1989}
\begin{center}
\begin{tabular}{rrrrrrrrrrrr}
\tableline
MJD & 10 & 20 & 34 & 65 & 100 & 150 & 250 & 350 \\
\tableline
  5480 & 252 & 244 & 237 & 230 & 227 & 224 & 222 & 221 \\
  5520 & 248 & 240 & 234 & 227 & 224 & 221 & 219 & 218 \\
  5560 & 246 & 238 & 232 & 225 & 222 & 219 & 217 & 216 \\
  5600 & 244 & 237 & 231 & 224 & 221 & 218 & 216 & 215 \\
  5640 & 243 & 235 & 229 & 223 & 219 & 217 & 215 & 214 \\
  5680 & 243 & 235 & 228 & 222 & 218 & 216 & 214 & 213 \\
  5720 & 243 & 234 & 228 & 221 & 218 & 215 & 213 & 212 \\
\end{tabular} \end{center} \end{table}

\begin{table}[p]
\caption{Brightness Temperature of Mars:1989-1995}
\begin{center}
\begin{tabular}{rrrrrrrrrrrr}
\tableline
MJD & 10 & 20 & 34 & 65 & 100 & 150 & 250 & 350 \\
\tableline
  7680 & 235 & 225 & 218 & 211 & 208 & 205 & 203 & 202 \\
  7720 & 238 & 229 & 222 & 215 & 212 & 209 & 207 & 206 \\
  7760 & 241 & 233 & 226 & 219 & 216 & 214 & 211 & 210 \\
  7800 & 245 & 236 & 230 & 223 & 220 & 217 & 215 & 214 \\
  7840 & 248 & 240 & 233 & 227 & 223 & 221 & 219 & 218 \\
  7880 & 250 & 242 & 236 & 229 & 225 & 223 & 220 & 219 \\
  7920 & 253 & 245 & 238 & 231 & 227 & 225 & 222 & 221 \\
\end{tabular} \end{center} \end{table}

\clearpage

\begin{table}[p]
\caption{Brightness Temperature of Mars:1995-2001}
\begin{center}
\begin{tabular}{rrrrrrrrrrrr}
\tableline
MJD & 10 & 20 & 34 & 65 & 100 & 150 & 250 & 350 \\
\tableline
  9880 & 233 & 222 & 215 & 208 & 205 & 202 & 200 & 199 \\
  9920 & 237 & 226 & 220 & 213 & 209 & 207 & 204 & 203 \\
  9960 & 242 & 232 & 225 & 218 & 215 & 212 & 210 & 209 \\
 10000 & 248 & 238 & 231 & 224 & 221 & 218 & 216 & 215 \\
 10040 & 255 & 245 & 238 & 231 & 227 & 224 & 222 & 221 \\
 10080 & 262 & 252 & 245 & 237 & 234 & 231 & 229 & 228 \\
 10120 & 267 & 258 & 251 & 243 & 240 & 237 & 235 & 234 \\
\end{tabular} \end{center} \end{table}
\begin{table}[p]
\caption{Brightness Temperature of Mars:2001-2007}
\begin{center}
\begin{tabular}{rrrrrrrrrrrr}
\tableline
MJD & 10 & 20 & 34 & 65 & 100 & 150 & 250 & 350 \\
\tableline
 12080 & 256 & 247 & 241 & 234 & 230 & 227 & 225 & 224 \\
 12120 & 249 & 237 & 229 & 222 & 218 & 215 & 213 & 212 \\
 12160 & 246 & 232 & 223 & 215 & 212 & 209 & 207 & 206 \\
 12200 & 247 & 233 & 225 & 217 & 213 & 210 & 208 & 206 \\
 12240 & 249 & 236 & 228 & 221 & 217 & 215 & 212 & 211 \\
 12280 & 248 & 237 & 229 & 222 & 218 & 216 & 213 & 212 \\
 12320 & 246 & 236 & 228 & 221 & 218 & 215 & 213 & 212 \\
\end{tabular} \end{center} \end{table}
\begin{table}[p]
\caption{Brightness Temperature of Mars:2007-2013}
\begin{center}
\begin{tabular}{rrrrrrrrrrrr}
\tableline
MJD & 10 & 20 & 34 & 65 & 100 & 150 & 250 & 350 \\
\tableline
 14280 & 265 & 255 & 248 & 241 & 238 & 235 & 233 & 232 \\
 14320 & 260 & 250 & 243 & 235 & 231 & 228 & 226 & 225 \\
 14360 & 255 & 245 & 237 & 229 & 225 & 223 & 221 & 219 \\
 14400 & 252 & 242 & 235 & 228 & 225 & 222 & 220 & 219 \\
 14440 & 250 & 242 & 236 & 229 & 225 & 223 & 221 & 220 \\
 14480 & 240 & 231 & 224 & 217 & 214 & 211 & 209 & 208 \\
 14520 & 231 & 220 & 212 & 205 & 202 & 199 & 197 & 196 \\
\end{tabular} \end{center} \end{table}
\begin{table}[p]
\caption{Brightness Temperature of Mars:2013-2019}
\begin{center}
\begin{tabular}{rrrrrrrrrrrr}
\tableline
MJD & 10 & 20 & 34 & 65 & 100 & 150 & 250 & 350 \\
\tableline
 16480 & 253 & 244 & 238 & 231 & 227 & 225 & 222 & 221 \\
 16520 & 248 & 240 & 234 & 227 & 224 & 221 & 219 & 218 \\
 16560 & 245 & 237 & 231 & 225 & 221 & 219 & 217 & 216 \\
 16600 & 243 & 235 & 229 & 223 & 219 & 217 & 215 & 214 \\
 16640 & 242 & 234 & 228 & 221 & 218 & 215 & 213 & 212 \\
 16680 & 242 & 234 & 228 & 221 & 218 & 215 & 213 & 212 \\
 16720 & 245 & 237 & 230 & 224 & 220 & 218 & 215 & 214 \\
\end{tabular} \end{center} \end{table}
\begin{table}[p]
\caption{Brightness Temperature of Mars:2019-2025}
\begin{center}
\begin{tabular}{rrrrrrrrrrrr}
\tableline
MJD & 10 & 20 & 34 & 65 & 100 & 150 & 250 & 350 \\
\tableline
 18680 & 240 & 231 & 224 & 218 & 214 & 212 & 210 & 209 \\
 18720 & 242 & 234 & 227 & 221 & 217 & 215 & 213 & 212 \\
 18760 & 244 & 236 & 230 & 223 & 220 & 217 & 215 & 214 \\
 18800 & 246 & 238 & 232 & 225 & 222 & 219 & 217 & 216 \\
 18840 & 248 & 239 & 233 & 226 & 222 & 220 & 218 & 217 \\
 18880 & 249 & 241 & 234 & 226 & 223 & 220 & 218 & 217 \\
 18920 & 252 & 243 & 236 & 229 & 225 & 222 & 220 & 219 \\
\end{tabular} \end{center} \end{table}
\begin{table}[p]
\caption{Brightness Temperature of Mars:2025-2031}
\begin{center}
\begin{tabular}{rrrrrrrrrrrr}
\tableline
MJD & 10 & 20 & 34 & 65 & 100 & 150 & 250 & 350 \\
\tableline
 20880 & 237 & 227 & 220 & 214 & 210 & 208 & 205 & 204 \\
 20920 & 243 & 233 & 226 & 219 & 216 & 213 & 211 & 210 \\
 20960 & 248 & 239 & 232 & 225 & 222 & 219 & 217 & 216 \\
 21000 & 254 & 245 & 238 & 231 & 228 & 225 & 223 & 222 \\
 21040 & 260 & 251 & 245 & 238 & 234 & 231 & 229 & 228 \\
 21080 & 266 & 258 & 251 & 244 & 240 & 238 & 235 & 234 \\
 21120 & 270 & 261 & 254 & 247 & 243 & 241 & 239 & 238 \\
\end{tabular} \end{center} \end{table}
\begin{table}[p]
\caption{Brightness Temperature of Mars:2031-2037}
\begin{center}
\begin{tabular}{rrrrrrrrrrrr}
\tableline
MJD & 10 & 20 & 34 & 65 & 100 & 150 & 250 & 350 \\
\tableline
 23080 & 241 & 228 & 221 & 214 & 210 & 208 & 205 & 204 \\
 23120 & 243 & 230 & 222 & 214 & 210 & 208 & 206 & 205 \\
 23160 & 248 & 234 & 226 & 218 & 214 & 211 & 209 & 208 \\
 23200 & 253 & 240 & 232 & 224 & 220 & 218 & 215 & 214 \\
 23240 & 255 & 243 & 236 & 228 & 225 & 222 & 220 & 219 \\
 23280 & 254 & 243 & 236 & 229 & 225 & 223 & 220 & 219 \\
 23320 & 251 & 241 & 235 & 228 & 224 & 222 & 219 & 218 \\
\end{tabular} \end{center} \end{table}
\begin{table}[p]
\caption{Brightness Temperature of Mars:2037-2043}
\begin{center}
\begin{tabular}{rrrrrrrrrrrr}
\tableline
MJD & 10 & 20 & 34 & 65 & 100 & 150 & 250 & 350 \\
\tableline
 25280 & 262 & 252 & 245 & 237 & 234 & 231 & 229 & 228 \\
 25320 & 261 & 250 & 243 & 235 & 231 & 229 & 226 & 225 \\
 25360 & 260 & 251 & 244 & 237 & 233 & 231 & 228 & 227 \\
 25400 & 250 & 240 & 234 & 227 & 223 & 221 & 218 & 217 \\
 25440 & 237 & 226 & 219 & 212 & 208 & 206 & 204 & 203 \\
 25480 & 231 & 219 & 212 & 205 & 201 & 199 & 197 & 196 \\
 25520 & 229 & 218 & 211 & 204 & 200 & 198 & 195 & 194 \\
\end{tabular} \end{center} \end{table}
\begin{table}[p]
\caption{Brightness Temperature of Mars:2043-2049}
\begin{center}
\begin{tabular}{rrrrrrrrrrrr}
\tableline
MJD & 10 & 20 & 34 & 65 & 100 & 150 & 250 & 350 \\
\tableline
 27480 & 251 & 242 & 236 & 229 & 225 & 223 & 220 & 219 \\
 27520 & 246 & 238 & 232 & 225 & 222 & 219 & 217 & 216 \\
 27560 & 243 & 236 & 229 & 223 & 219 & 217 & 215 & 214 \\
 27600 & 242 & 234 & 228 & 222 & 218 & 216 & 213 & 212 \\
 27640 & 243 & 235 & 229 & 222 & 219 & 216 & 214 & 213 \\
 27680 & 243 & 235 & 229 & 222 & 219 & 216 & 214 & 213 \\
 27720 & 237 & 227 & 220 & 214 & 210 & 208 & 205 & 204 \\
\end{tabular} \end{center} \end{table}
\begin{table}[p]
\caption{Brightness Temperature of Mars:2049-2055}
\begin{center}
\begin{tabular}{rrrrrrrrrrrr}
\tableline
MJD & 10 & 20 & 34 & 65 & 100 & 150 & 250 & 350 \\
\tableline
 29680 & 242 & 235 & 229 & 222 & 219 & 216 & 214 & 213 \\
 29720 & 244 & 236 & 230 & 223 & 220 & 217 & 215 & 214 \\
 29760 & 245 & 237 & 231 & 224 & 221 & 218 & 216 & 215 \\
 29800 & 245 & 237 & 231 & 224 & 220 & 218 & 216 & 215 \\
 29840 & 246 & 237 & 230 & 223 & 219 & 217 & 214 & 213 \\
 29880 & 248 & 238 & 231 & 224 & 220 & 218 & 215 & 214 \\
 29920 & 252 & 242 & 235 & 228 & 224 & 222 & 220 & 219 \\
\end{tabular} \end{center} \end{table}
\begin{table}[p]
\caption{Brightness Temperature of Mars:2055-2061}
\begin{center}
\begin{tabular}{rrrrrrrrrrrr}
\tableline
MJD & 10 & 20 & 34 & 65 & 100 & 150 & 250 & 350 \\
\tableline
 31880 & 243 & 234 & 227 & 220 & 217 & 214 & 212 & 211 \\
 31920 & 248 & 239 & 232 & 226 & 222 & 220 & 217 & 216 \\
 31960 & 253 & 244 & 237 & 230 & 227 & 224 & 222 & 221 \\
 32000 & 258 & 249 & 243 & 236 & 232 & 230 & 227 & 226 \\
 32040 & 263 & 255 & 248 & 241 & 238 & 235 & 233 & 232 \\
 32080 & 268 & 259 & 253 & 245 & 242 & 239 & 237 & 236 \\
 32120 & 269 & 260 & 253 & 246 & 243 & 240 & 238 & 237 \\
\end{tabular} \end{center} \end{table}
\begin{table}[p]
\caption{Brightness Temperature of Mars:2061-2067}
\begin{center}
\begin{tabular}{rrrrrrrrrrrr}
\tableline
MJD & 10 & 20 & 34 & 65 & 100 & 150 & 250 & 350 \\
\tableline
 34080 & 244 & 232 & 224 & 217 & 213 & 211 & 208 & 207 \\
 34120 & 250 & 237 & 229 & 222 & 218 & 215 & 213 & 212 \\
 34160 & 256 & 244 & 236 & 228 & 224 & 222 & 219 & 218 \\
 34200 & 260 & 249 & 242 & 234 & 231 & 228 & 226 & 225 \\
 34240 & 261 & 250 & 243 & 236 & 232 & 230 & 227 & 226 \\
 34280 & 258 & 249 & 242 & 235 & 231 & 229 & 227 & 226 \\
 34320 & 254 & 245 & 239 & 232 & 228 & 226 & 223 & 222 \\
\end{tabular} \end{center} \end{table}
\begin{table}[p]
\caption{Brightness Temperature of Mars:2067-2073}
\begin{center}
\begin{tabular}{rrrrrrrrrrrr}
\tableline
MJD & 10 & 20 & 34 & 65 & 100 & 150 & 250 & 350 \\
\tableline
 36280 & 268 & 259 & 253 & 245 & 242 & 239 & 237 & 236 \\
 36320 & 257 & 246 & 239 & 231 & 228 & 225 & 223 & 222 \\
 36360 & 244 & 233 & 226 & 219 & 215 & 213 & 210 & 209 \\
 36400 & 237 & 226 & 219 & 212 & 209 & 206 & 204 & 203 \\
 36440 & 234 & 223 & 216 & 209 & 205 & 203 & 201 & 200 \\
 36480 & 233 & 222 & 215 & 208 & 204 & 202 & 200 & 199 \\
 36520 & 234 & 224 & 217 & 210 & 206 & 204 & 201 & 200 \\
\end{tabular} \end{center} \end{table}
\begin{table}[p]
\caption{Brightness Temperature of Mars:2073-2079}
\begin{center}
\begin{tabular}{rrrrrrrrrrrr}
\tableline
MJD & 10 & 20 & 34 & 65 & 100 & 150 & 250 & 350 \\
\tableline
 38480 & 248 & 239 & 232 & 226 & 222 & 220 & 217 & 216 \\
 38520 & 245 & 236 & 230 & 223 & 220 & 217 & 215 & 214 \\
 38560 & 243 & 236 & 230 & 223 & 220 & 217 & 215 & 214 \\
 38600 & 243 & 236 & 230 & 223 & 220 & 217 & 215 & 214 \\
 38640 & 237 & 228 & 221 & 214 & 210 & 208 & 205 & 204 \\
 38680 & 231 & 221 & 214 & 207 & 203 & 201 & 198 & 197 \\
 38720 & 232 & 221 & 214 & 207 & 204 & 201 & 199 & 198 \\
\end{tabular} \end{center} \end{table}
\begin{table}[p]
\caption{Brightness Temperature of Mars:2079-2085}
\begin{center}
\begin{tabular}{rrrrrrrrrrrr}
\tableline
MJD & 10 & 20 & 34 & 65 & 100 & 150 & 250 & 350 \\
\tableline
 40680 & 243 & 236 & 230 & 223 & 220 & 217 & 215 & 214 \\
 40720 & 244 & 236 & 230 & 223 & 220 & 217 & 215 & 214 \\
 40760 & 244 & 236 & 229 & 223 & 219 & 217 & 214 & 213 \\
 40800 & 244 & 235 & 228 & 221 & 218 & 215 & 213 & 212 \\
 40840 & 246 & 236 & 229 & 222 & 218 & 215 & 213 & 212 \\
 40880 & 251 & 241 & 235 & 227 & 224 & 221 & 219 & 218 \\
 40920 & 257 & 248 & 242 & 235 & 231 & 229 & 227 & 226 \\
\end{tabular} \end{center} \end{table}
\begin{table}[p]
\caption{Brightness Temperature of Mars:2085-2091}
\begin{center}
\begin{tabular}{rrrrrrrrrrrr}
\tableline
MJD & 10 & 20 & 34 & 65 & 100 & 150 & 250 & 350 \\
\tableline
 42880 & 247 & 238 & 232 & 225 & 222 & 219 & 217 & 216 \\
 42920 & 251 & 242 & 236 & 229 & 226 & 223 & 221 & 220 \\
 42960 & 254 & 246 & 239 & 232 & 229 & 226 & 224 & 223 \\
 43000 & 259 & 250 & 244 & 237 & 233 & 231 & 229 & 227 \\
 43040 & 263 & 255 & 248 & 241 & 238 & 235 & 233 & 232 \\
 43080 & 266 & 257 & 250 & 243 & 240 & 237 & 235 & 234 \\
 43120 & 266 & 257 & 250 & 243 & 239 & 236 & 234 & 233 \\
\end{tabular} \end{center} \end{table}
\begin{table}[p]
\caption{Brightness Temperature of Mars:2091-2097}
\begin{center}
\begin{tabular}{rrrrrrrrrrrr}
\tableline
MJD & 10 & 20 & 34 & 65 & 100 & 150 & 250 & 350 \\
\tableline
 45080 & 251 & 240 & 232 & 225 & 221 & 219 & 217 & 215 \\
 45120 & 258 & 247 & 239 & 232 & 228 & 225 & 223 & 222 \\
 45160 & 263 & 253 & 246 & 238 & 235 & 232 & 230 & 229 \\
 45200 & 266 & 256 & 249 & 242 & 238 & 236 & 233 & 232 \\
 45240 & 264 & 255 & 248 & 241 & 238 & 235 & 233 & 232 \\
 45280 & 260 & 251 & 245 & 238 & 234 & 232 & 229 & 228 \\
 45320 & 254 & 246 & 239 & 232 & 229 & 226 & 224 & 223 \\
\end{tabular} \end{center} \end{table}
\begin{table}[p]
\caption{Brightness Temperature of Mars:2097-2103}
\begin{center}
\begin{tabular}{rrrrrrrrrrrr}
\tableline
MJD & 10 & 20 & 34 & 65 & 100 & 150 & 250 & 350 \\
\tableline
 47280 & 249 & 236 & 228 & 220 & 217 & 214 & 212 & 211 \\
 47320 & 245 & 233 & 226 & 218 & 215 & 212 & 210 & 209 \\
 47360 & 242 & 231 & 224 & 217 & 213 & 211 & 208 & 207 \\
 47400 & 240 & 229 & 222 & 215 & 211 & 209 & 207 & 206 \\
 47440 & 238 & 227 & 220 & 214 & 210 & 208 & 205 & 204 \\
 47480 & 237 & 228 & 221 & 214 & 210 & 208 & 206 & 205 \\
 47520 & 239 & 230 & 223 & 216 & 213 & 210 & 208 & 207 \\
\end{tabular} \end{center} \end{table}

\end{document}